\documentclass[reprint,superscriptaddress,showpacs,amsmath, amssymb,aps,pra]{revtex4-2}
\usepackage{graphicx}
\usepackage{dcolumn}
\usepackage{bm}
\usepackage{subfigure}
\usepackage{hyperref}
\hypersetup{colorlinks=true, citecolor=blue, urlcolor=blue, linkcolor=blue}
\usepackage[ruled]{algorithm2e}

\usepackage{epsfig,graphicx}
\usepackage{amsmath}
\usepackage{amssymb}
\usepackage{mathrsfs}
\usepackage{url}
\usepackage{amsthm}
\usepackage{bm}
\usepackage{hyperref}
\usepackage{amsmath, bm}
\usepackage{float}
\usepackage{epstopdf}
\usepackage{tabularx}
\usepackage{cases}
\usepackage{color}

\hypersetup{colorlinks=true, citecolor=blue, urlcolor=blue, linkcolor=blue}
\newtheorem{theorem}{Theorem}
\newtheorem{lemma}{Lemma}

\newtheorem{exmp}{Example}

\newtheorem{definition}{Definition}
\usepackage{color}

\begin{document}
\title{Tight upper bound for the maximal expectation value of the $N$-partite generalized Svetlichny operator}
\author{Youwang Xiao}
\affiliation{College of the Science, China University of Petroleum, 266580, Qingdao, China}
\author{Zong Wang}
\affiliation{School of Mathematical Sciences, Shanghai Jiao Tong University, 200240, Shanghai,  China}
\author{Wen-Na Zhao}
\affiliation{College of the Science, China University of Petroleum, 266580, Qingdao, China}
\author{Ming Li}
\email{liming@upc.edu.cn.}
\affiliation{College of the Science, China University of Petroleum, 266580, Qingdao, China}

\date{\today}
\begin{abstract}
 Genuine multipartite non-locality is not only of fundamental interest but also serves as an important resource for quantum information theory. We consider the $N$-partite scenario and provide an analytical upper bound on the maximal expectation value of the generalized Svetlichny  inequality achieved by an arbitrary $N$-qubit system.  
  Furthermore, the constraints on quantum states for which the upper bound is tight are also presented and illustrated by noisy generalized Greenberger-Horne-Zeilinger (GHZ) states. Especially, the new techniques proposed to derive the upper bound allow more insights into the structure of the generalized Svetlichny  operator and enable us to systematically investigate the relevant properties. As an operational approach, the variation of the correlation matrix we defined makes it more convenient to search for suitable unit vectors that satisfy the tightness conditions. Finally, our results give feasible experimental implementations in detecting the genuine multipartite non-locality and can potentially be applied to other quantum information processing tasks.
\end{abstract}
\maketitle
\smallskip
\section{Introduction}
Correlations that can be described in terms of local hidden variables necessarily satisfy a set of linear constraints known as the Bell inequalities \cite{ref1},  showing that there is an upper limit for the
 correlations predicted by local realism theory. Nevertheless, a violation of this limit can be observed from the quantum correlations generated by performing local measurements on entangled particles \cite{ref2,ref3,ref4}, which indicates that the predictions of quantum theory are incompatible with local realism theory and we thus refer to this  striking phenomenon as quantum non-locality \cite{ref5}. Non-local correlations witnessed by the violation of Bell inequalities has been identified as an extremely  useful quantum resource for many applications in quantum-information theory such as
the reduction of communication complexity \cite{ref6}, secure quantum key distribution \cite{ref7} and device-independent random number generation \cite{ref8,ref9}.

Compared to the bipartite cases, the multipartite correlations exhibit much richer and more complex structures in the multipartite scenarios \cite{ref10,ref11,ref12}. As the strongest notation of multipartite non-locality, genuine multipartite non-locality (GMNL) is one of the most fundamental non-classical features of multipartite systems and has attracted considerable attention \cite{ref13,ref14,ref15,ref16,ref17}. Initially, Svetlichny \cite{ref3} provided a Bell-type inequality to detect genuine tripartite non-locality and  later the Svetlichny inequality has been generalized to arbitrary parties \cite{ref18,ref19} and arbitrary dimensions \cite{ref20}. Nevertheless, the task of detection and characterization of GMNL is demanding, as the complexity of the possible states of the systems and sets of correlations grows exponentially with the number of parties involved in multipartite Bell scenario. Multipartite Bell-type inequalities are an effective approach to describe multipartite correlations and provide insight into the rich structure of multipartite scenarios and thus much effort has been devoted to this research with a desire to gain a better understanding of GMNL \cite{ref17,ref21,ref22,ref23}. Furthermore, the significant advantage of GMNL  lies in certifying the presence of genuine multipartite entanglement  in a device-independent way by observing the violation of some genuine multipartite Bell-type inequalities. For progress in the experimental aspects of multipartite entanglement, see references \cite{ref24} and \cite{ref25}.

To quantitatively analyze the genuine tripartite non-locality existing in the general three-qubit states, a method is developed to compute the maximal violation  of the Svetlichny inequality \cite{ref3} and a tight upper bound is obtained in Ref. \cite{ref21}. Recently, the first upper
bound on the maximal violation of the  Mermin-Ardehali-Belinskii-Klyshko (MABK) inequality achievable by an arbitrary $N$-qubit state has been proven in Ref.\,\,\cite{ref26}, where the techniques arouse our interest in a deeper exploration of the $N$-partite generalized  Svetlichny (GS) inequality \cite{ref19}. In particular, the research for explicit analytical expressions for the maximal violation of the bipartite \cite{ref27,ref28} and multipartite Bell inequalities for arbitrary quantum states is of great relevance, not only contributing to  capture the deviation of quantum correlations from classical ones but also serving as a key component of secure cryptographic protocols \cite{ref26,ref29}.

 In this work, we provide an analytical upper bound on the quantum expectation value of the generalized Svetlichny operators for an arbitrary $N$-qubit state for the two cases $N$ even and $N$ odd, respectively,  with the intention of gaining  a better understanding of GMNL by considering the generalized  Svetlichny inequality.
 Our analysis gives more insights into the structure of the generalized Svetlichny inequality, notable that the techniques  of our derivation of the upper bound are different from those in Ref.\,\,\cite{ref26}, where with our definition, the variation of the correlation matrix can, to a degree(for instance, the violation of $N$-partite GS inequality by a particular class of target states), make it more convenient to search for suitable unit vectors that satisfy the tightness conditions. In addition, our $N$-partite bound is tight for certain classes of states depending on the corresponding tightness conditions, which is illustrated by noisy quantum states and extends the known results \cite{ref21} for three-qubit states. Finally, for the non-locality witnessed by the Svetlichny inequality our results may be  potentially useful in the context of sharing  genuine non-locality \cite{ref30}, the non-locality breaking property of  channels \cite{ref31} and device-independent secret sharing \cite{ref32} in the multipartite scenarios.
\section{THE $N$-PARTITE GS INEQUALITY.}
Contrary to those in bipartite systems, quantum
states in multipartite systems can be not only local or
non-local  but also genuinely non-local, which can be
revealed by violation of genuine multipartite Bell inequalities.  We consider a Bell scenario in which $N$ spatially separated observers, Alice$_{1}$, ..., Alice$_{N}$, share an $N$-qubit quantum state and each of them can measure the binary observable $A_{x_{i}}^{(i)} (i=1,...,N), x_{i}\in \left \{ 0,1 \right \} $ with outcome $a_{i} =\pm 1$ locally on their shared part of the system.  Let $P(a_{1}\cdots a_{N}|A_{x_{1}}^{(1)}\cdots A_{x_{N}}^{(N)})$ denote the jointly conditional probabilities (or correlations) where Alice$_{i}$ measures her system by  $A_{x_{i}}^{(i)}$ with outcome $a_{i}$. Following the definition of \cite{ref18,ref19}, the correlations are called genuinely multipartite non-local if they cannot be decomposed into the hybrid local-non-local form
\begin{align}\label{hylnon}
&\qquad P(a_{1}\cdots a_{N}|A_{x_{1}}^{(1)}\cdots A_{x_{N}}^{(N)})\notag\\
&\qquad=\sum_{i_{1}\cdots i_{k}}q_{i_{1}\cdots i_{k}} \int d\lambda \, \rho_{i_{1}\cdots i_{k}}(\lambda )\notag\\ &\qquad\quad\times P_{i_{1}\cdots i_{k}}(a_{i_{1}}\cdots a_{i_{k}}|A_{x_{i_{1}}}^{(i_{1})}\cdots A_{x_{i_{k}}}^{(i_{k})},\lambda )\notag\\
&\qquad\quad\times P_{{i_{k+1}\cdots i_{N}}}(a_{i_{k+1}}\cdots a_{i_{N}}|A_{x_{i_{k+1}}}^{(i_{k+1})}\cdots A_{x_{i_{N}}}^{(i_{N})},\lambda),
\end{align}
where $\left \{ i_{1},..., i_{k} \right \}\bigcup  \left \{ i_{k+1},..., i_{N} \right \}=\left \{ 1,...,N \right \} $, the sum takes into account the different bipartitions of the parties and $q_{i_{1}\cdots i_{k}}$ denotes  the probability that allows arbitrary correlations among parties ${\rm{Alice}}_{i_{1}},..., {\rm{Alice}}_{i_{k}} $ and ${\rm{Alice}}_{i_{k+1}},..., {\rm{Alice}}_{i_{N}} $.

In particular, the $N$-partite generalized Svetlichny inequality was derived under the assumption of allowing arbitrarily strong correlations inside each  subsystem but no correlation between different subsystems  in Ref. \cite{ref18,ref19} and  the violation of the GS inequality is a signature of genuine multipartite non-locality. Specifically, we now introduce the generalized Svetlichny operator defined as
\begin{align}
S_{N}^{\pm }=\sum_{x}\nu _{t(x)}^{\pm } A _{x_{1}}^{(1) }\otimes  \cdots \otimes    A _{x_{N}}^{(N) },
\end{align}
where  $x=(x_{1},...,x_{N})$, $t(x)$  is the number of times element 1 appears in  $x$, $\nu _{k}^{\pm}$ is the sign function given by $\nu _{k}^{\pm}=(-1)^{k(k\pm1)/2} $ and $A _{x_{i}}^{(i)}$ for $x_{i}=0,1$ are the binary observables
of Alice$_{i}$. These sequences of $\nu _{t(x)}^{\pm}$ have period four with cycles
$(1, -1, -1, 1)$ for positive  sign and $(1, 1, -1, -1)$ for negative sign. Note that the GS operator can also be defined by recursion
\begin{align}\label{2recur}
S_{N}^{\pm }=S_{N-1}^{\pm }A_{0}^{(N)}\mp S_{N-1}^{\mp }A_{1}^{(N)},
\end{align}
However, it should be emphasized that $S_{N}^{+}$ can be obtained from $S_{N}^{-}$ by applying the mapping $A_{0}^{N}\to -A_{1} ^{N}$ and $A_{1}^{N}\to A_{0}^{N}$ and thus is one of its equivalent forms.
For any $N$-qubit state $\rho$ with the hybrid local-non-local form (\ref{hylnon}), the GS inequality  then can be expressed as
\begin{align}
|\left \langle S_{N}^{\pm } \right \rangle _{\rho }|=\left | {\rm{Tr}}(\rho S_{N}^{\pm } )  \right | \le 2^{N-1},
\end{align}
where a violation of the bound $2^{N-1}$ implies that $\rho$ features genuine multipartite non-locality. In quantum mechanics, the GS inequality is maximally violated up to a value of $2^{N-1}\sqrt2$ by the general GHZ state
\begin{align}
\left | {\rm{GHZ}}^{n}   \right \rangle=\frac{1}{\sqrt2} (\left | 0 \right \rangle^{\otimes n}+\left | 1 \right \rangle^{\otimes n}  ).
\end{align}

\section{PRELIMINARIES}
In this section, we introduce some definitions and lemmas to prepare the ground for proving the main results.

It is well known that each single-qubit quantum state has a Bloch sphere representation, parametrized by the following relation
\begin{align}
\rho =\frac{1}{2}\sum_{\mu=0}^{3}\Lambda _{\mu}\sigma _{\mu},
\end{align}
where $\Lambda _{0}=1,\sigma _{0}=I_{2}$ is the $2\times2$ identity operator and $\sigma _{i}$ are Pauli operators for $i=1,2,3$. Similarly and slightly more generally, one can
generalize this unique expression to an arbitrary $N$-qubit quantum state via tensor product of Pauli matrices as
\begin{align}\label{S2}
\rho =\frac{1}{2^{N} }\sum_{\mu _{1}\dots \mu _{N} =0}^{3} \Lambda _{\mu _{1}\dots \mu _{N}}\sigma _{\mu_{1} } \otimes \dots \otimes \sigma _{\mu _{N} }.
\end{align}

Then, the correlation matrix of an $N$-qubit state can be expressed as follows.
\begin{definition}
 We  define $M_{\rho }^{N}$ the correlation matrix of an $N$-qubit state $\rho$ with the elements $\left [ M_{\rho }^{N} \right ]_{r_{j_{1}\dots j_{N-2 }},c_{j_{N-1} j_{N } }}$  given by
\begin{align}\label{TM}
\left [ M_{\rho }^{N} \right ] _{r_{j_{1}\dots j_{N-2 } },c_{j_{N-1} j_{N } } }&=\Lambda _{j_{1}\dots j_{N}}\notag\\
&={\rm{Tr}}\left [\rho\, \sigma _{j_{1}} \otimes \cdots \otimes  \sigma _{j_{N}}   \right ],
\end{align}
where
\begin{align}
&r_{j_{1}\dots j_{N-2 }}=1+\sum_{i=1}^{N-2} 3^{N-2-i}(j_{i}-1), \\
&c_{j_{N-1} j_{N } } =3(j_{N-1}-1)+j_{
N},
\end{align}
and $j_{1},...,j_{N}\in \left \{ 1,2,3 \right \} $.
\end{definition}
\begin{lemma}
For any $m\times n$ real matrix $Q$, vectors  $\vec{x}\in \mathbb{R}^{m}$ and $\vec{y}\in \mathbb{R}^{n}$, let $\left \| \vec{v}  \right \| $ be the Euclidean norm of vector $\vec{v}$ and then we have
\begin{align}
|\vec{x}^{T}  Q\,\vec{y} | \le \sigma _{max} \left \| \vec{x}  \right \|\left \| \vec{y}  \right \|,
\end{align}
where  $\sigma _{max}$ is the maximum singular value of the matrix $Q$ and $T$ denotes the matrix transposition. The equality
holds when $\vec{x}$ and $\vec{y}$ are the corresponding singular vectors of $Q$ with respect to $\sigma _{max}$.
\end{lemma}
For the detailed proof, we refer readers
to \cite{ref21}.
\begin{lemma}
Let $w(\bm{x})$ be the Hamming weight of a bit string $\bm{x}=(x_{1},..,x_{N}  )$ defined by
\begin{align}
 w(\bm{x}) =\left | \left \{ i|x_{i}=1,1\le i\le N  \right \}  \right |,
\end{align}
then, the number of binary bit  strings with an odd Hamming weight equals those with an even Hamming weight.
\end{lemma}
See the proof in Appendix A.
\begin{definition}
For  ${\bm{m_{1}}}, {\bm{m_{2}}}\in \left \{ 0,1 \right \}^{N}$, let $\left | {\bm{m_{1}}}\bigcap {\bm{m_{2}}}   \right | $ denotes the number of elements in two equal-length binary bit strings that differ in corresponding positions. $\left | {\bm{m_{1}}}\bigcap {\bm{m_{2}}}   \right |_{j_{1}\cdots j_{k} } $ specifically indicates that the elements in positions $j_{1},..., j_{k}$ are different with $1\le j_{1}< \cdots<  j_{k}\le N$.
\end{definition}
\begin{lemma}
Assuming $1\le\left | {\bm{m_{1}}}\bigcap {\bm{m_{2}}}   \right |=k\le N-1 $ for bit strings ${\bm{m_{1}}}, {\bm{m_{2}}}\in \left \{ 0,1 \right \}^{N}$, we divide ${\bm{m_{1}}}$ into ${\bm{m_{1}^{d}}}$  and ${\bm{m_{s}}}$ and  ${\bm{m_{2}}}$ into ${\bm{m_{2}^{d}}}$  and ${\bm{m_{s}}}$ satisfying $\bm{m_{1}^{d}}\oplus  \bm{m_{2}^{d}}={\bm{e_{k}}}$,  respectively, where $\oplus$ is the binary modulo 2 addition and ${\bm{e_{k}}}$ is a vector of length $k$ with all elements being one. Then the following identities hold
\begin{align}
&\left\lfloor  \frac{w({\bm{m_{1}^{d}}})+w({\bm{m_{s}}})}{2}  \right\rfloor+\left\lfloor  \frac{w({\bm{m_{2}^{d}}})+w({\bm{m_{s}}})}{2}  \right\rfloor\notag\\
&\qquad=\left\{\begin{matrix}
 \frac{k-1}{2}+ w({\bm{m_{s}}}) & k\; {\rm{odd}}\\
 \frac{k}{2}+ 2 \left\lfloor w({\bm{m_{s}}})/2 \right\rfloor & k \; {\rm{even}}, w({\bm{m_{1}^{d}}})\; {\rm{even}}\\
 \frac{k-2}{2}+ 2\left\lceil w({\bm{m_{s}}})/2 \right\rceil &k \; {\rm{even}}, w({\bm{m_{1}^{d}}})\; {\rm{odd}},
\end{matrix}\right.\\
&\left\lfloor  \frac{w({\bm{m_{1}^{d}}})+w({\bm{m_{s}}})}{2}  \right\rfloor+\left\lceil  \frac{w({\bm{m_{2}^{d}}})+w({\bm{m_{s}}})}{2}  \right\rceil\notag\\
&\qquad=\left\{\begin{matrix}
 \frac{k}{2}+ w({\bm{m_{s}}}) & k\; {\rm{even}}\\
 \frac{k+1}{2}+ 2 \left\lfloor w({\bm{m_{s}}})/2 \right\rfloor & k \; {\rm{odd}}, w({\bm{m_{1}^{d}}})\; {\rm{even}}\\
 \frac{k-1}{2}+ 2\left\lceil w({\bm{m_{s}}})/2 \right\rceil &k \; {\rm{odd}}, w({\bm{m_{1}^{d}}})\; {\rm{odd}},
\end{matrix}\right.\label{UL}
\end{align}
where  $\left\lceil x \right\rceil$ and $\left\lfloor x \right\rfloor $  are the ceiling and floor functions, returning the closest integer that is greater than or equal to the function argument $x$ and the closest integer that is less than or equal to the function argument $x$, respectively.
\end{lemma}
See the proof in Appendix B. This lemma will be particularly useful towards the  subsequent discussion of the properties of high dimensional vectors  strongly constrained by their tensor product structure.

\section{Tight UPPER BOUND ON GS INEQUALITY VIOLATION.}
As stated by the following theorem, the first primary result of this paper is a tight upper bound on the maximum quantum value of the GS operator for an arbitrary $N$-qubit state shared by the parties in the case of odd $N$.
\begin{theorem}
The maximum quantum value $\mathcal{GS}_{\rho}$ of the $N$-partite GS inequality for an arbitrary $N$-qubit state $\rho$  with odd $N$ satisfies
\begin{align}\label{Theo1}
\mathcal{GS}_{\rho}\equiv \max_{A _{x_{i}}^{(i)}} |\left \langle S_{N}^{\pm } \right \rangle _{\rho }|\le 2^{\frac{N+1}{2} }\sigma _{max},
\end{align}
where $A _{x_{i}}^{(i)}=\vec{a}_{x_{i}}^{i}\cdot \vec{\sigma }    $ is  projective measurement with unit vector $\vec{a}_{x_{i} }^{i}=({a}_{x_{i},1 }^{i},{a}_{x_{i},2 }^{i},{a}_{x_{i},3 }^{i})\in \mathbb{R}^{3}$ (for $i=1,...,N$) and $ \vec{\sigma } =(\sigma _{1},\sigma _{2},\sigma _{3} )$, and where $\sigma _{max} $ is  the maximum singular value of the correlation matrix $M_{\rho }^{N}$  as follows from Definition 1.
\end{theorem}
Formally, we now proceed to prove the Theorem 1.
\begin{proof}
First of all, we present  analytical expressions for the $N$-partite GS operator $S_{N}^{\pm } $. In particular, we distinguish the case $S_{N}^{-} $:
\begin{align}
S_{N}^{-} =\sum_{\bm{x}\in \left \{ 0,1 \right \}^{N} }(-1)^{\left\lfloor w(\bm{x})/2\right\rfloor }\underset{i=1}{\overset{N}{\bigotimes }}    A_{x_{i} }^{(i)},
\end{align}
and the case $S_{N}^{+} $:
\begin{align}
S_{N}^{+} =\sum_{\bm{x}\in \left \{ 0,1 \right \}^{N} }(-1)^{\left\lceil w(\bm{x})/2\right\rceil }\underset{i=1}{\overset{N}{\bigotimes }}    A_{x_{i} }^{(i)},
\end{align}

For the sake of equivalence of $S_{N}^{\pm }$, we now specify the expression of the GS inequality $\left \langle S_{N}^{-}  \right \rangle _{\rho }$ for an arbitrary $N$-qubit quantum state $\rho$ when $N$ is odd. Note that the GS inequality is not equivalent to the MABK inequality at this point. Using the recursive form (\ref{2recur}), we obtain
\begin{align}\label{proof11}
S_{N}^{- }&=S_{N-1}^{- }A_{0}^{(N)}+S_{N-1}^{+ }A_{1}^{(N)}\notag\\
&=(S_{N-2}^{- }A_{0}^{(N-1)}+S_{N-2}^{+ }A_{1}^{(N-1)})A_{0}^{(N)}\notag\\
&\qquad+(S_{N-2}^{+ }A_{0}^{(N-1)}-S_{N-2}^{-}A_{1}^{(N-1)})A_{1}^{(N)}\notag\\
&=S_{N-2}^{- }(A_{0}^{(N-1)}A_{0}^{(N)}-A_{1}^{(N-1)}A_{1}^{(N)})\notag\\
&\qquad+S_{N-2}^{+}(A_{0}^{(N-1)}A_{1}^{(N)}+A_{1}^{(N-1)}A_{0}^{(N)}),
\end{align}
Then we can derive an explicit expression for the GS expectation value as follows:
\begin{align}\label{proof12}
&\left \langle S_{N}^{-}  \right \rangle _{\rho }={\rm{Tr}}\left [ S_{N}^{-} \rho  \right ]\notag\\
&=\sum_{j_{1},...,j_{N}=1  }^{3} \left [ \sum_{x\in \left \{ 0,1 \right \}^{N-2}  }\sum_{y\in E^{2}  }(-1)^{\left\lfloor w({\bm{x}})/2\right\rfloor+\left\lfloor w({\bm{y}})/2\right\rfloor  }\right.\notag\\
&\left. \qquad a_{x_{1},j_{1}  }^{1}\dots a_{x_{N-2},j_{N-2}  }^{N-2} \Lambda _{j _{1}\dots j _{N}}a_{y_{1},j_{N-1}  }^{N-1}a_{y_{2},j_{N}  }^{N}\right.\notag\\
&\left.\qquad\qquad\quad+ \sum_{x\in \left \{ 0,1 \right \}^{N-2}  }\sum_{y\in O^{2}  }(-1)^{\left\lceil w({\bm{x}})/2\right\rceil +\left\lfloor w({\bm{y}})/2\right\rfloor  }\right.\notag\\
&\left. \qquad a_{x_{1},j_{1}  }^{1}\dots a_{x_{N-2},j_{N-2}  }^{N-2} \Lambda _{j _{1}\dots j _{N}}a_{y_{1},j_{N-1}  }^{N-1}a_{y_{2},j_{N}  }^{N}  \right ],
\end{align}
where the sets $E^{2}$ and $O^{2}$ are defined  in the following way
\begin{align}
&E^{2} =\left \{ {\bm{y}}\in \left \{ 0,1 \right \}^{2}| w({\bm{y}}) \;{\rm{mod}} 2=0  \right \},\\
&O^{2} =\left \{ {\bm{y}}\in \left \{ 0,1 \right \}^{2}| w({\bm{y}}) \;{\rm{mod}} 2=1 \right \}.
\end{align}

By defining the vectors,
\begin{align}
&\vec{v}_{0}=  \sum_{{\bm{x}}\in \left \{ 0,1 \right \}^{N-2}   }(-1)^{\left\lfloor w({\bm{x}})/2\right\rfloor }\bigotimes _{i=1}^{N-2}\vec{a}_{x_{i}}^{i},\label{v1}\\
&\vec{v}_{1}=  \sum_{{\bm{x}}\in \left \{ 0,1 \right \}^{N-2}   }(-1)^{\left\lceil w({\bm{x}})/2\right\rceil  }\bigotimes _{i=1}^{N-2}\vec{a}_{x_{i}}^{i},\label{v2}\\
&\vec{u}_{0}=  \sum_{{\bm{y}}\in {E}^{2}   }(-1)^{\left\lfloor w({\bm{y}})/2\right\rfloor }\bigotimes _{i=N-1}^{N}\vec{a}_{y_{i}}^{i},\label{v3}\\
&\vec{u}_{1}=  \sum_{{\bm{y}}\in {O}^{2}   }(-1)^{\left\lfloor w({\bm{y}})/2\right\rfloor }\bigotimes _{i=N-1}^{N}\vec{a}_{y_{i}}^{i},\label{v4}
\end{align}
it immediately follows from the Definition 1 of the correlation matrix of an $N$-qubit state that
\begin{align}\label{GS}
\left \langle S_{N}^{-}  \right \rangle _{\rho }=\vec{v}_{0}^{T}M_{\rho }^{N}\vec{\mu }_{0}+ \vec{v}_{1}^{T}M_{\rho }^{N}\vec{\mu }_{1}.
\end{align}

The following important properties satisfied by these vectors in Eqs. (\ref{v1})-(\ref{v4}) (see Appendix C for the detailed proof) contribute significantly to the derivation of valid upper bounds on $\left \langle S_{N}^{-}  \right \rangle _{\rho }$.
\begin{align}
&{\mathbf{Prop.1:}}\qquad\left \| \vec{v}_{0}  \right \|^{2}+\left \| \vec{v}_{1}  \right \|^{2}=2^{N-1},\\
&{\mathbf{Prop.2:}}\qquad\left \| \vec{u}_{0}  \right \|^{2}+\left \| \vec{u}_{1}  \right \|^{2}=4,\\
&{\mathbf{Prop.3:}}\qquad\vec{v}_{0}\cdot  \vec{v}_{1}=0,\\
&{\mathbf{Prop.4:}}\qquad\vec{u}_{0}\cdot   \vec{u}_{1}=0.
\end{align}
As a  result of the properties above, we can formulate the vectors $\vec{v}_{k}$ and $\vec{u}_{k}$ $(k=0,1)$ as follows:
\begin{subequations}
\begin{align}
&\vec{v}_{0}=2^{\frac{N-1}{2} }{\rm{cos}}\,\gamma _{\mathcal{O} }\,  \hat{v}_{0} \label{1:sub1},\\
&\vec{v}_{1}=2^{\frac{N-1}{2} }{\rm{sin}}\,\gamma _{\mathcal{O} }\,  \hat{v}_{1} \label{1:sub2},
\end{align}
\end{subequations}
and
\begin{subequations}
\begin{align}
&\vec{u}_{0}=2\,{\rm{cos}}\,\beta  _{\mathcal{O} }\,  \hat{u}_{0} \label{2:sub1},\\
&\vec{u}_{1}=2\,{\rm{sin}}\,\beta  _{\mathcal{O} }\,  \hat{u}_{1} \label{2:sub2},
\end{align}
\end{subequations}
where $\hat{v}_{k}$ and $\hat{u}_{k}$ are unit vectors in the directions of $\vec{v}_{k}$ and $\vec{u}_{k}$, respectively, and where $\gamma _{\mathcal{O} },
\beta  _{\mathcal{O} }$ are  angles between 0 and $\frac{\pi }{2} $.
Using these expressions Eqs.\,\,(\ref{1:sub1}), (\ref{1:sub2}), (\ref{2:sub1}) and (\ref{2:sub2}), we can recast the GS expectation value, Eqs. (\ref{GS}), as follows:
\begin{align}\label{SGS}
\left \langle S_{N}^{-}  \right \rangle _{\rho }&=2^{\frac{N-1}{2} }{\rm{cos}}\gamma _{\mathcal{O} } \, \hat{v}_{0}^{T}\cdot M_{\rho }^{N}\cdot 2\,{\rm{cos}}\beta  _{\mathcal{O} } \, \hat{u}_{0}\notag\\
&\quad+2^{\frac{N-1}{2} }{\rm{sin}}\,\gamma _{\mathcal{O} } \, \hat{v}_{1}^{T}\cdot M_{\rho }^{N}\cdot 2\,{\rm{sin}}\,\beta  _{\mathcal{O} }\, \hat{u}_{1}\notag\\
&=2^{\frac{N+1}{2} }\left [  {\rm{cos}}\,\gamma _{\mathcal{O} }\, {\rm{cos}}\, \beta_{\mathcal{O} }\, \hat{v}_{0}^{T}\cdot M_{\rho }^{N}\cdot \hat{u}_{0}\right.\notag\\
&\left.\quad\qquad\quad+{\rm{sin}}\,\gamma _{\mathcal{O} }\, {\rm{sin}}\, \beta _{\mathcal{O} }\, \hat{v}_{1}^{T}\cdot M_{\rho }^{N}\cdot \hat{u}_{1}\right ],
\end{align}

The maximum quantum value $\mathcal{GS}_{\rho}$ of the $N$-partite GS
inequality is then obtained by maximizing Eq.\,\,(\ref{SGS}) over all measurement directions $\vec{a}_{0}^{i}$ and $\vec{a}_{1}^{i}$ (for $i=1,...,N $) of the parties. A feasible upper bound on expectation value is thus yielded by
\begin{align}\label{UPbound}
|\left \langle S_{N}^{-}  \right \rangle _{\rho }| &\le 2^{\frac{N+1}{2} }\sigma _{max}( {\rm{cos}}\,\gamma _{\mathcal{O} }\, {\rm{cos}}\, \beta _{\mathcal{O} }\, +{\rm{sin}}\,\gamma _{\mathcal{O} }\, {\rm{sin}}\,\beta _{\mathcal{O} } )\notag\\
&=2^{\frac{N+1}{2} }\sigma _{max}\,{\rm{cos}}(\gamma _{\mathcal{O} }-\beta _{\mathcal{O} })\notag\\
&\le2^{\frac{N+1}{2} }\sigma _{max}.
\end{align}
where we have used Lemma 1 for the first inequality and trigonometric identity for the second inequality and where $\sigma _{max}$ is the largest singular value of the matrix $M_{\rho }^{N}$.

This concludes the proof.
\end{proof}
\subsection*{1.\,\,Tightness conditions for odd $N$}
In the following, we derive the conditions for the case $N$ odd under which the upper bound on the maximum GS inequality violation obtained in Theorem 1 is tight.
From Lemma 1, it follows that the first inequality in Eq.\,\,(\ref{UPbound}) is saturated if the degeneracy of $\sigma _{max}$ is more than 1,  and corresponding to $\sigma _{max}$, there exist appropriate singular vectors in the form of  $\vec{v}_{k}$ and $\vec{u}_{k}$ defined in  (\ref{v1})$-$(\ref{v4}), respectively. Furthermore, we can fix the measuring directions of the parties by specifying $\gamma _{\mathcal{O} }=\beta _{\mathcal{O} }$ to make the second inequality in Eq.\,\,(\ref{UPbound}) become an equality. It is straightforward to check that the tightness conditions ensure every equality sign holds, showing that the upper bound is attained.

The following theorem, presented as the second main result, is the analogue of the previous one (\ref{Theo1}), where the tight upper bound for even $N$ derived by our method is further discussed and compared with their counterparts shown in \cite{ref26}.

\begin{theorem}
The maximum quantum value $\mathcal{GS}_{\rho}$ of the $N$-partite GS inequality for an arbitrary $N$-qubit state $\rho$  with even $N$ satisfies
\begin{align}
\mathcal{GS}_{\rho}\equiv \max_{A _{x_{i}}^{(i)}} |\left \langle S_{N}^{\pm } \right \rangle _{\rho }|\le 2^{\frac{N}{2} }\sqrt{\lambda _{1}+\lambda _{2}},
\end{align}
where $\lambda _{1}$ and $\lambda _{2}$ are the the squares of the first and second largest singular values of the matrix $M_{\rho }^{N}$ as follows from Definition 1, respectively.
\end{theorem}
\begin{proof}
In fact, vectors $\vec{v}_{0}$ and $\vec{v}_{1}$ in Eq.\,\,(\ref{v1})and Eq.\,\,(\ref{v2}) satisfy the property $\left \| \vec{v}_{0}  \right \|^{2}=\left \| \vec{v}_{1}  \right \|^{2}=2^{N-2} $ for even $N$ cases. We now proceed to prove it, where we replace $\vec{v}_{k}$ ($k=0,1$)  with $\vec{v}_{k}^{N} $ for the convenience of distinguishing between odd and even cases. With the same arrangement of correlation matrix given in Definition
1, note that the following identities exist for even $N$:
\begin{align}
\vec{v}_{0}^{N+1}=\vec{v}_{0}^{N}\otimes \vec{a}_{0}^{N+1}+\vec{v}_{1}^{N}\otimes \vec{a}_{1}^{N+1},\notag\\
\vec{v}_{1}^{N+1}=\vec{v}_{1}^{N}\otimes \vec{a}_{0}^{N+1}-\vec{v}_{0}^{N}\otimes \vec{a}_{1}^{N+1}.
\end{align}
By employing properties 1 and 3,  the conclusion can be readily verified. Therefore, vectors  $\vec{v}_{k}$ and $\vec{u}_{k}(k=0,1)$, Eqs. (\ref{v1})$-$(\ref{v4}), can be represented as
\begin{subequations}
\begin{align}
&\vec{v}_{k}=2^{\frac{N-2}{2} }\hat{v}_{k}, \label{111:sub1}\\
&\vec{u}_{0}=2\,{\rm{cos}}\,\alpha_{\mathcal{E} }\,  \hat{u}_{0}, \label{222:sub1}\\
&\vec{u}_{1}=2\,{\rm{sin}}\,\alpha_{\mathcal{E} }\,  \hat{u}_{1}, \label{222:sub2}
\end{align}
\end{subequations}
where $\hat{v}_{k}$ and $\hat{u}_{k}$ are unit vectors in the directions of $\vec{v}_{k}$ and $\vec{u}_{k}$, respectively, and  $\alpha _{\mathcal{E} }$ is  the angle between 0 and $\frac{\pi }{2} $ for the even cases.
Using these expressions, we obtain the GS expectation value for even $N$:
\begin{align}\label{GSEven}
&\left \langle S_{N}^{-}  \right \rangle _{\rho }=2^{\frac{N-2}{2} } \, \hat{v}_{0}^{T}\cdot M_{\rho }^{N}\cdot 2\,{\rm{cos}}\,\alpha_{\mathcal{E} } \, \hat{u}_{0}\notag\\
&\qquad\qquad+2^{\frac{N-2}{2} }\, \hat{v}_{1}^{T}\cdot M_{\rho }^{N}\cdot 2\,{\rm{sin}}\, \alpha_{\mathcal{E} }\, \hat{u}_{1}\notag\\
&=2^{\frac{N}{2} }\left [  {\rm{cos}}\, \alpha_{\mathcal{E} }\, \hat{v}_{0}^{T}\cdot M_{\rho }^{N}\cdot \hat{u}_{0}+ {\rm{sin}}\, \alpha_{\mathcal{E} }\, \hat{v}_{1}^{T}\cdot M_{\rho }^{N}\cdot \hat{u}_{1}\right ],
\end{align}
A feasible upper bound on expectation value is thus yielded by
\begin{align}\label{UPboundEven}
|\left \langle S_{N}^{-}  \right \rangle _{\rho }| &\le  2^{\frac{N}{2} }\left [  {\rm{cos}} \, \alpha_{\mathcal{E} }\, \left \|  M_{\rho }^{N}\cdot \hat{u}_{0} \right \| + {\rm{sin}} \, \alpha_{\mathcal{E} }\, \left \|  M_{\rho }^{N}\cdot \hat{u}_{1} \right \| \right ]\notag\\
&\le2^{\frac{N}{2} }\sqrt{\left \|  M_{\rho }^{N}\cdot \hat{u}_{0} \right \|^{2}+\left \|  M_{\rho }^{N}\cdot \hat{u}_{1} \right \|^{2}}\notag\\
&\le 2^{\frac{N}{2} }\sqrt{\lambda _{1}+\lambda _{2}}.
\end{align}
where we use the fact $\vec{x}^{T}\vec{y}\le \left \| \vec{x} \right \|\cdot   \left \| \vec{y} \right \| $, Cauchy-Schwartz inequality and the result of Lemma 2 given in Ref. \cite{ref26} accordingly to achieve the upper bound and where $\lambda _{1}$ and $\lambda _{2}$ are the the squares of the largest and second-to-the-largest singular values of $M_{\rho }^{N}$, respectively.

This concludes the proof.
\end{proof}
\subsection*{2.\,\, Tightness conditions for even $N$}
So far, except for a slight modification by the constant $\frac{1}{2^{(N-2)/2} } $, we have presented the same upper bound for even $N$ as in Ref. \cite{ref26}, where  only the case $N/2$ even is explicitly derived in the proof. Next, we will consider the conditions under which tightness is maintained. We first fix the directions of $\hat{v}_{0}$ and $\hat{v}_{1}$ to those of $M_{\rho }^{N}\cdot \hat{u}_{0}$ and $M_{\rho }^{N}\cdot \hat{u}_{1}$, respectively, as follows:
\begin{align}\label{ss1}
\hat{v}_{k}= \frac{ M_{\rho }^{N}\cdot \hat{u}_{k} }{\left \| M_{\rho }^{N}\cdot \hat{u}_{k} \right \|},
\end{align}
so that the first line of Eq.\,\,(\ref{UPboundEven}) becomes an equation. Additionally, if one specify angle $\alpha_{\mathcal{E} }$ between $\hat{u}_{0}$ and $\hat{u}_{1}$  such that
\begin{align}\label{ss2}
{\rm{tan}}\,\alpha_{\mathcal{E} } =\frac{\left \| M_{\rho }^{N}\cdot \hat{u}_{1} \right \| }{\left \| M_{\rho }^{N}\cdot \hat{u}_{0} \right \|},
\end{align}
 the second inequality in Eq.\,\,(\ref{UPboundEven}) is saturated. Finally, the saturation of the last inequality requires choosing $\hat{u}_{k}$ that satisfy the relation
 \begin{align}\label{ss3}
 (M_{\rho }^{N})^{T} M_{\rho }^{N}\hat{u}_{k}=\lambda _{k}\hat{u}_{k}.
 \end{align}
 where $\lambda _{k}$ $ (k=0,1)$ are the two largest singular values of $M_{\rho }^{N}$. It can be verified that once unit vectors $\vec{a}_{x_{i} }^{i}$ ($i=1,...,N$) take the form of Eqs.\,\,(\ref{v1})$-$(\ref{v4}) and satisfy tightness conditions of Eqs.\,\,(\ref{ss1})$-$(\ref{ss3}), the upper bound in Eq.\,\,(\ref{UPboundEven}) is achieved.

 The case in which our tightness holds is slightly different from that in \cite{ref26}, mainly in terms of the tensor-product structure of the vectors we defined, Eqs. (\ref{v1})$-$(\ref{v4}), as a result of the alternative definition of the correlation matrix of an $N$-qubit state given by Definition 1. The following examples show that this variation of the correlation matrix can, to a degree(for instance, the violation of $N$-partite GS inequality by a particular class of target states), make it more convenient to search for suitable unit vectors that satisfy the tightness conditions. Indeed, as an operational approach, it is possible to further extend the idea from expression (\ref{proof11}) that we can constantly redefine the correlation matrix  associated with an $N$-qubit in a particular order to determine the desired unit vectors.

\section{TIGHTNESS ANALYSIS FOR NOISY GENERALIZED GHZ STATES.}
Subsequently, we will mainly explore noisy $N$-qubit generalized GHZ states as examples to discuss the violation of GS inequality in detail, considering that these states have specific entanglement properties which arise from their inherent symmetry and render them candidates for information processing protocols.
\begin{exmp}
For $N=4$, consider the mixture of the white noise and the four-qubit  generalized GHZ states, given by
\begin{equation}\label{Ex2}
\rho_{g_{4}} =p |\psi _{g_{4}}\rangle\langle\psi _{g_{4}}|+\frac{1-p}{16}I_{4},
\end{equation}
where $I_{4}$ is the identity matrix on $(\mathbb{C}^{2})^{\otimes 4} $, $|\psi _{g_{4}}\rangle= {\rm{cos}}\,\theta_{1} |0000\rangle+{\rm{sin}}\,\theta_{1} |1111\rangle$ and $0\le p\le 1$.
\end{exmp}
By definition, the correlation matrix $M_{\rho }^{N}$ of $\rho_{g_{4}}$ has the form
\begin{align}\label{CoreT}
M_{\rho }^{N}=p\begin{pmatrix}
a  & 0 & 0 & 0 & -a & 0 & 0 & 0 & 0\\
0  & -a & 0 & -a & 0 & 0 &  0&0  &0 \\
0  & 0 & 0 &0  & 0 & 0 & 0 & 0 &0 \\
0  & -a & 0 & -a & 0 & 0 & 0 & 0 & 0\\
 -a & 0 & 0 & 0 & a & 0 & 0 & 0 &0 \\
 0 & 0 & 0 &  0& 0 &  0&0  &  0& 0\\
 0 & 0 & 0 &  0& 0 &  0& 0 & 0 & 0\\
 0 &0  &  0& 0 &  0&  0& 0 & 0 &0 \\
0  & 0 &  0&0  & 0 & 0 & 0 & 0 &1
\end{pmatrix},
\end{align}
where $a =2{\rm{cos}}\,\theta_{1} \,{\rm{sin}}\,\theta_{1} $. The nonzero singular values corresponding to Eq.\,\,(\ref{CoreT}) are $\sigma _{1}=\sigma _{2}=2p|{\rm{sin}}\,2\theta _{1}|$ and $\sigma _{3}=p$.
 If $|{\rm{sin}}\,2\theta _{1}|\ge \frac{1}{2} $, that is, the matrix $M_{\rho }^{N}$ is degenerate in its largest singular value, we achieve ${\rm{max}} |\left \langle S_{N}^{- } \right \rangle _{\rho_{g_{4}} }|=8\sqrt{2} p|{\rm{sin}}\,2\theta _{1}|$, with settings
 \begin{align}\label{Dire1}
&\vec{a}_{0} =(-\frac{1}{\sqrt{2} } ,-\frac{1}{\sqrt{2} } ,0)^{T},\qquad \vec{a}_{1} =(\frac{1}{\sqrt{2} } ,-\frac{1}{\sqrt{2} } ,0)^{T},\notag\\
& \vec{b}_{0} =(1,0,0)^{T},\qquad \vec{b}_{1} =(0,1,0)^{T},\notag\\
&\vec{c}_{0} =(1,0,0)^{T},\qquad \vec{c}_{1} =(0,1,0)^{T},\notag\\
&\vec{d}_{0} =(0,1,0)^{T},\qquad \vec{d}_{1} =(-1,0,0),^{T}
\end{align}
 such that
\begin{align}
  &\vec{v}_{0}= \vec{a}_{0}\otimes ( \vec{b}_{0}+ \vec{b}_{1})+ \vec{a}_{1}\otimes ( \vec{b}_{0}- \vec{b}_{1}),\notag\\
  &\vec{v}_{1}= \vec{a}_{0}\otimes ( \vec{b}_{0}- \vec{b}_{1})- \vec{a}_{1}\otimes ( \vec{b}_{0}+ \vec{b}_{1}),
\end{align}
and
\begin{align}
 &\vec{u}_{0}=\vec{c}_{0}\otimes \vec{d}_{0}-\vec{c}_{1}\otimes \vec{d}_{1},\notag\\
 &\vec{u}_{1}=\vec{c}_{0}\otimes \vec{d}_{1}+\vec{c}_{1}\otimes\vec{d}_{0},
\end{align}
 to ensure the upper bound is saturated for $\rho_{g_{4}}$, where the state $\rho_{g_{4}}$  will violate the GS inequality as long as  $\sqrt{2}p|{\rm{sin}}\,2\theta _{1}| >1$.  For the other case $|{\rm{sin}}\,2\theta _{1}|< \frac{1}{2} $, there exist no unit vectors $ \vec{c}_{k}$ and $ \vec{d}_{k}$ ($k=0,1$) that satisfy the relation
 \begin{align}
 \vec{c}_{0}\otimes \vec{d}_{0}-\vec{c}_{1}\otimes \vec{d}_{1}=2\sqrt{\frac{1}{1+2|{\rm{sin}}2\theta _{1}|} }\vec{t}_{0},\notag\\
  \vec{c}_{0}\otimes \vec{d}_{1}+\vec{c}_{1}\otimes \vec{d}_{0}=2\sqrt{\frac{2|{\rm{sin}}2\theta _{1}|}{1+2|{\rm{sin}}2\theta _{1}|} }\vec{t}_{1}.
 \end{align}
 where $\vec{t}_{0}=(0,0,0,0,0,0,0,0,1)^{T}$ is the singular vector associated with maximum singular value $p$ and similarly for $\vec{t}_{1}=(0,1,0,1,0,0,0,0,1)^{T}$ or $\vec{t}_{1}=(-1,0,0,0,1,0,0,0,0)^{T}$, with corresponding singular value $2p|{\rm{sin}}\,2\theta _{1}|$, which means  the tightness conditions no longer hold. Nevertheless, it does not prevent us from utilizing the upper bound to deduce that  $\rho_{g_{4}}$ cannot exhibit genuine multipartite non-locality in the range $|{\rm{sin}}\,2\theta _{1}|< \frac{1}{2} $, due to the fact that ${\rm{max}} |\left \langle S_{N}^{- } \right \rangle _{\rho_{g_{4}} }|\le 4p\sqrt{1+4{\rm{sin}}^{2}\,2\theta _{1} }\le4\sqrt{2} < 8$.
\begin{exmp}
For $N=5$, consider the mixture of the white noise and the five-qubit  generalized GHZ  states, given by
\begin{equation}\label{Ex1}
\rho_{g_{5}} =p |\psi _{g_{5}}\rangle\langle\psi _{g_{5}}|+\frac{1-p}{32}I_{5},
\end{equation}
where $I_{5}$ is the identity matrix on $(\mathbb{C}^{2})^{\otimes 5} $, $|\psi _{g_{5}}\rangle= {\rm{cos}}\,\theta_{2} |00000\rangle+{\rm{sin}}\,\theta_{2} |11111\rangle$ and $0\le p\le 1$.
\end{exmp}
Similarly, the correlation matrix $M_{\rho }^{N}$ of $\rho_{g_{5}}$ turns out to be
\begin{align}
M_{\rho }^{N}=p\begin{pmatrix}
 A & B & O\\
 B & -A & O\\
 O & O & O\\
 B & -A & O\\
 -A & -B & O\\
 O & O & O\\
 O & O & O\\
 O & O & O\\
 O & O & C
\end{pmatrix},
\end{align}
where submatrixes $A, B, C$ are given by
\begin{align}
A=\begin{pmatrix}
\lambda   & 0 & 0\\
 0 & -\lambda  & 0\\
0  & 0 & 0
\end{pmatrix},\,\, B=\begin{pmatrix}
0   & -\lambda & 0\\
 -\lambda & 0  & 0\\
0  & 0 & 0
\end{pmatrix},\,\, C=\begin{pmatrix}
0   & 0 & 0\\
0 & 0  & 0\\
0  & 0 & u
\end{pmatrix},
\end{align}
with $\lambda =2{\rm{cos}}\,\theta_{2} \,{\rm{sin}}\,\theta_{2} $ and $u={\rm{cos}}^{2}\,\theta_{2} -{\rm{sin}}^{2}\,\theta_{2} $ and submatrix $O$ is a zero matrix.

Consequently, one can easily obtain that the nontrivial singular values of the matrix $M_{\rho }^{N}$ are
$\sigma _{1}=\sigma _{2}=2\sqrt{2 }p|{\rm{sin}}\,2\theta_{2}  |   $ and $\sigma _{3}=p\sqrt{1-{\rm{sin}}^{2}\,2\theta_{2}  }  $. Therefore, to satisfy  the required tightness conditions for $\rho_{g_{5}}$,  we can simply assume $\sigma _{max}=2\sqrt{2 }p|{\rm{sin}}\,2\theta_{2}  | $ by  suitably choosing   parameter $\theta_{2} $ (${\rm{tan}}^{2}\,2\theta_{2} \ge \frac{1}{8}$). Then we can set
\begin{align}\label{Dire}
&\vec{a}_{0} =(-\frac{1}{\sqrt{2} } ,-\frac{1}{\sqrt{2} } ,0)^{T},\qquad \vec{a}_{1} =(\frac{1}{\sqrt{2} } ,-\frac{1}{\sqrt{2} } ,0)^{T},\notag\\
& \vec{b}_{0} =(1,0,0)^{T},\qquad \vec{b}_{1} =(0,1,0)^{T},\notag\\
&\vec{c}_{0} =(1,0,0)^{T},\qquad \vec{c}_{1} =(0,1,0)^{T},\notag\\
&\vec{d}_{0} =(1,0,0)^{T},\qquad \vec{d}_{1} =(0,1,0)^{T},\notag\\
&\vec{e}_{0} =(0,1,0)^{T},\qquad \vec{e}_{1} =(-1,0,0)^{T},
\end{align}
as the measurement directions of corresponding parties  to guarantee that $\vec{v}_{k}$ and $\vec{u}_{k}$ decomposed as
\begin{align}
 &\vec{v}_{0}=\left [  \vec{a}_{0}\otimes ( \vec{b}_{0}+ \vec{b}_{1})+ \vec{a}_{1}\otimes ( \vec{b}_{0}- \vec{b}_{1}) \right ]\otimes \vec{c}_{0}\notag\\
&\quad\quad+\left [  \vec{a}_{0}\otimes ( \vec{b}_{0}- \vec{b}_{1})- \vec{a}_{1}\otimes ( \vec{b}_{0}+ \vec{b}_{1}) \right ]\otimes \vec{c}_{1},\notag\\
& \vec{v}_{1}=\left [  \vec{a}_{0}\otimes ( \vec{b}_{0}- \vec{b}_{1})- \vec{a}_{1}\otimes ( \vec{b}_{0}+ \vec{b}_{1}) \right ]\otimes \vec{c}_{0}\notag\\
 &\quad\quad-\left [  \vec{a}_{0}\otimes ( \vec{b}_{0}+ \vec{b}_{1})+ \vec{a}_{1}\otimes ( \vec{b}_{0}- \vec{b}_{1}) \right ]\otimes \vec{c}_{1},
\end{align}
and
\begin{align}
 &\vec{u}_{0}=\vec{d}_{0}\otimes \vec{e}_{0}-\vec{d}_{1}\otimes \vec{e}_{1},\notag\\
 &\vec{u}_{1}=\vec{d}_{0}\otimes \vec{e}_{1}+\vec{d}_{1}\otimes\vec{e}_{0}.
\end{align}
become the singular vectors for $\sigma _{max}$. In this case, one can  verify that $\gamma _{\mathcal{O} }=\beta _{\mathcal{O} }=\frac{\pi }{4} $, and that each of the inequalities in Eq. (\ref{UPbound}) becomes equal, which means that the upper bound in Theorem 1 is achieved for $\rho_{g_{5}}$ in Eq. (\ref{Ex1}).

Hence, we obtain the optimal value of GS operator ${\rm{max}}|\left \langle S_{N}^{- } \right \rangle _{\rho_{g_{5}} }|=16\sqrt{2} p|{\rm{sin}}\,2\theta _{2}|$, where the state $\rho_{g_{5}}$ in Eq.\,\,(\ref{Ex1}) can violate the GS inequality if and only if $\sqrt{2}p|{\rm{sin}}\,2\theta _{2}| >1$. It is not difficult to observe that the threshold value of $\theta_{2}$, above which  $\rho_{g_{5}}$ demonstrates genuine non-local correlations, is the same as in the case of $\rho_{g_{4}}$, that is, $|{\rm{sin}}\,2\theta _{2}|> \frac{1}{\sqrt{2}} $. Notice that the upper bound in Theorem 1 for $\rho_{g_{5}}$ is tight within the range ${\rm{tan}}^{2}\,2\theta_{2} \ge \frac{1}{8}$. The fact  ${\rm{max}} |\left \langle S_{N}^{- } \right \rangle _{\rho_{g_{5}} }|\le 8p\sqrt{1-{\rm{sin}}^{2}\,2\theta _{2} }< 16$ shows that $\rho_{g_{5}}$ can never violate the GS inequality if ${\rm{tan}}^{2}\,2\theta_{2} < \frac{1}{8}$. It is worth remarking that the range in which tightness holds for $\rho_{g_{5}}$ has increased as compared to the case of $N=3$ in Ref. \cite{ref21}, where ${\rm{tan}}^{2}\,2\theta_{2} \ge \frac{1}{2}$ is required (see Fig. 1).


The results of Eq.\,\,(\ref{Dire1}) and  Eq.\,\,(\ref{Dire}) suggest that we can restrict directions of spin measurements of all the parties  lie in the $x-y$ plane to achieve the maximal value of GS inequality for noisy generalized GHZ states. Thus providing explicit measurement settings that maximize the violation of the GS inequality, these examples illustrate the importance of our results, which may be of particular value to the experimental application focused on genuine multi-qubit non-locality for the implementation of certain quantum information processing tasks \cite{ref33}.

Based on the discussion above and the numerical verification, we can extend the conclusion of noisy four-qubit and five-qubit generalized GHZ states to  general $N$-qubit case, expressed as follows:

\emph{Corollary 1.} For the mixture of the white noise and the $N$-qubit generalized GHZ state $\rho_{g_{N}}=p|\psi _{g_{N}}\rangle\langle\psi _{g_{N}}| +\frac{1-p}{2^{N}}I_{N} $,
where $|\psi _{g_{N}}\rangle={\rm{cos}}\, \theta \left | 0\cdots  0  \right \rangle+{\rm{sin}}\, \theta \left | 1\cdots  1  \right \rangle$, the nontrivial singular values of $M_{\rho }^{N}$ given in Definition 1 are
\begin{widetext}
\begin{align}
\Sigma(M_{\rho }^{N})=\begin{cases}
 2^{\frac{N-2}{2}p |{\rm{sin}}\,2\theta | }, 2^{\frac{N-2}{2}p |{\rm{sin}}\,2\theta | }, p\sqrt{1-{\rm{sin}}^{2}\,2\theta }  & ,\text{ for odd N} \\
 2^{\frac{N-2}{2}p |{\rm{sin}}\,2\theta | }, 2^{\frac{N-2}{2}p |{\rm{sin}}\,2\theta | },p & ,\text{ for even N },
\end{cases}
\end{align}
\end{widetext}
Our upper bound preserves  tightness for $\rho_{g_{N}}$ in the range ${\rm{tan}}^{2}\,\theta\ge \frac{1}{2^{N-2}} $ for odd $N$ and ${\rm{sin}}^{2}\,\theta\ge \frac{1}{2^{N-2}} $ for even $N$, respectively. In conclusion, we obtain the following claim,
\begin{widetext}
\begin{align}
{\rm{max}}|\left \langle S_{N}^{- } \right \rangle _{\rho_{g_{N}} }|=\begin{cases}
 2^{N-1}\sqrt{2}p|{\rm{sin}}\,2\theta|  &  ,\,\text{for} \,\,  {\rm{tan}}^{2}\,\theta\ge \frac{1}{2^{N-2}}\, \text{and odd N } \\
 2^{N-1}\sqrt{2}p|{\rm{sin}}\,2\theta|  &  ,\,\text{for}\,\,  {\rm{sin}}^{2}\,\theta\ge \frac{1}{2^{N-2}} \, \text{and even N },
\end{cases}
\end{align}
\end{widetext}
which can be achieved by setting
\allowdisplaybreaks[2]
\begin{align}
&\vec{a}_{0}^{1} =(-\frac{1}{\sqrt{2} } ,-\frac{1}{\sqrt{2} } ,0)^{T},\qquad \vec{a}_{1}^{1} =(\frac{1}{\sqrt{2} } ,-\frac{1}{\sqrt{2} } ,0)^{T},\notag\\
& \vec{a}_{0}^{i} =(1,0,0)^{T},\qquad\qquad\quad\,\,\; \vec{a}_{1}^{i} =(0,1,0)^{T},\notag\\
&\vec{a}_{0}^{N} =(0,1,0)^{T},\qquad\qquad\quad\;
\vec{a}_{1}^{N} =(-1,0,0)^{T}.
\end{align}
with $i=2,...,N-1$.
\begin{figure}[http]
 \centering
   \includegraphics[width=7.5cm,height=6.4cm]{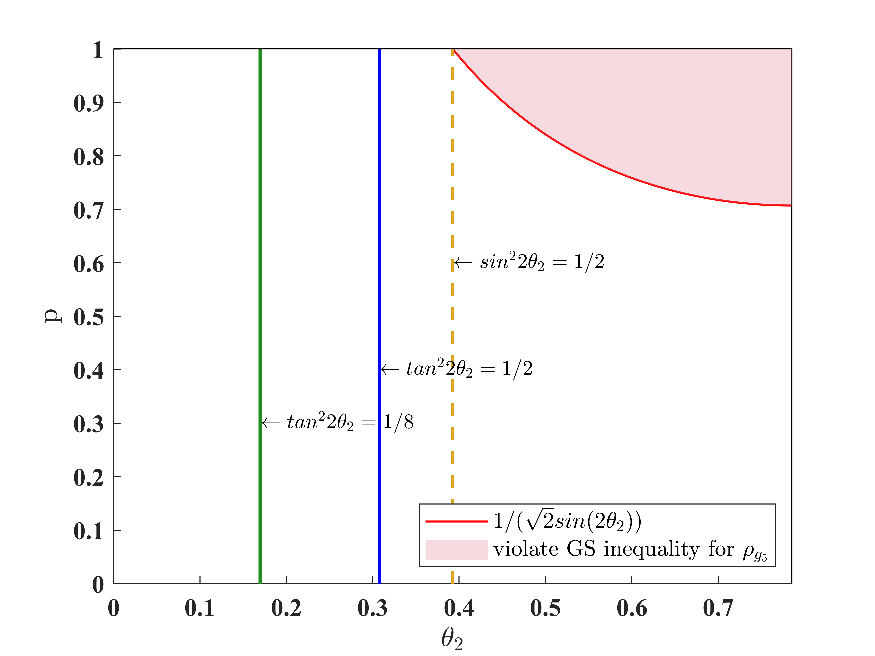}
  \caption{The upper bound is not tight for the quantum state $\rho_{g_{5}}$ that coincides with the range of parameters $\theta_{2}$ and $p$ represented by the blank area to the left of the green solid line. However, tightness holds for $\rho_{g_{5}}$ with ranges consistent with those represented by the region to the right of the green solid line, where the  states corresponding to the filled region above the red solid line can violate the generalized Svetlichny inequality and the threshold value of $\theta_{2}$ satisfies ${\rm{sin}}^{2}\,2\theta_{2}=1/{2} $ (orange dashed line). The range in which tightness holds has been extended compared to the case of $N=3$ (blue solid line). Note that the theorem presented in this paper is not only valid for pure states, but also works for mixed multi-qubit systems.}
\end{figure}
\section{Conclusions and Discussions}
The detection of genuine multipartite non-locality is a demanding task. To this end, we consider the $N$-partite scenario where the parties test a generalized Svetlichny Bell inequality with two measurement settings and two outcomes per party. We structurally provide an analytical upper bound on the maximal violation of the generalized Svetlichny  inequality attained by a given $N$-qubit state, which extends the known results for $N=3$ in Ref. \cite{ref21} to more qubits. Furthermore, we present a different technique to the derivation of the upper bound from that in Ref.\,\,\cite{ref24}, which offers new  perspectives for the generalized Svetlichny  operator and may inspire analogous analytical results for other Bell inequalities. The upper bound, independent of the measurement settings, is tight for certain classes of states where the tightness is explored through noisy quantum states given by  the mixture of the white noise and the $N$-qubit generalized GHZ states. Thus, our results can potentially serve as an effective criterion for the experimental detection of genuine multipartite non-locality and can be applied to various quantum information processing tasks. Finally, it would be also interesting to discover more  insights in the multipartite scenarios by utilizing the upper bound in the context of sharing genuine non-locality \cite{ref28}, the non-locality breaking
property of channels \cite{ref29} and device-independent secret sharing \cite{ref30}.

\section*{\bf Acknowledgments} This work is supported by the Shandong Provincial Natural Science Foundation for Quantum Science ZR2021LLZ002, the Fundamental Research Funds for the Central Universities No.22CX03005A.

\section*{\bf Data availability statement} All data generated or analyzed during this study are included in this published article.

\section*{APPENDIX A: PROOF OF LEMMA 2 }\label{app}
\begin{proof}
For the case of odd $N$, we have
\begin{align}
\sum_{x_{odd} }C_{N}^{x_{odd} }=  \sum_{x_{odd} }C_{N}^{N-x_{odd} }= \sum_{y_{even} }C_{N}^{y_{odd} }.
\end{align}
For the case of even $N$, we have
\begin{align}
\sum_{x_{odd} }C_{N}^{x_{odd} }&=  \sum_{x_{odd} }C_{N-1}^{x_{odd}-1 }+ \sum_{x_{odd} }C_{N-1}^{x_{odd} }\notag\\
&= \sum_{y_{even} }C_{N-1}^{y_{even} }+\sum_{x_{odd} }C_{N-1}^{x_{odd} }\notag\\
&=2^{N-1}\notag\\
&=\frac{1}{2}(\sum_{y_{even} }C_{N}^{y_{even} }+\sum_{x_{odd} }C_{N}^{x_{odd} }).
\end{align}
where $C_{N}^{k}=N!/[k!(N-k)!]$, $x_{odd}$ and $y_{even}$ denotes Hamming weight for odd numbers and even numbers, respectively, and we use the fact that $C_{N}^{k}=C_{N}^{N-k}$ and $C_{N}^{k}=C_{N-1}^{k-1}+C_{N-1}^{k}$. Therefore, we obtain $\sum_{x_{odd} }C_{N}^{x_{odd} }=\sum_{y_{even} }C_{N}^{y_{even} }$ for any $N$.
\end{proof}
\section*{APPENDIX B: PROOF OF LEMMA 3 }\label{app}
\begin{proof}
\emph{The former}.
Based on the following fact,
\begin{align}\label{L31}
\left\lfloor \frac{w({\bm{x}})+w({\bm{y}})}{2} \right\rfloor=\left\{\begin{matrix}
 \left\lfloor w({\bm{x}})/2 \right\rfloor+\left\lfloor w({\bm{y}})/2  \right\rfloor &, w({\bm{y}})\; {\rm{even}}  \\
  \left\lceil w({\bm{x}})/2  \right\rceil+\left\lfloor w({\bm{y}})/2  \right\rfloor &, w({\bm{y}})\; {\rm{odd}},
\end{matrix}\right.
\end{align}
for the case of odd $k$ and  odd $w({\bm{m_{1}^{d}}})$ we have
\begin{align}
&\left\lfloor  \frac{w({\bm{m_{1}^{d}}})+w({\bm{m_{s}}})}{2}  \right\rfloor+\left\lfloor  \frac{w({\bm{m_{2}^{d}}})+w({\bm{m_{s}}})}{2}  \right\rfloor\notag\\
&=\left\lfloor \frac{w({\bm{m_{1}^{d}}})}{2}  \right\rfloor+\left\lceil \frac{w({\bm{m_{s}}})}{2} \right\rceil+\left\lfloor \frac{w({\bm{m_{2}^{d}}})}{2} \right\rfloor+\left\lfloor  \frac{w({\bm{m_{s}}})}{2} \right\rfloor\notag\\
&=\frac{(w({\bm{m_{1}^{d}}})-1)}{2} +\frac{w({\bm{m_{2}^{d}}})}{2} +w({\bm{m_{s}}})\notag\\
&=\frac{k-1}{2} +w({\bm{m_{s}}}).
\end{align}
where we utilize $w({\bm{m_{1}^{d}}})+w({\bm{m_{2}^{d}}})=k$ and the fact that  $\left\lfloor  x/2 \right\rfloor+\left\lceil x/2 \right\rceil=x$ together with the fact that $\left\lfloor x/2 \right\rfloor=(x-1)/2$ for odd $x$. A similar procedure applied to the case of odd $k$ and even $w({\bm{m_{1}^{d}}})$ can  result  in the same result and thus we omit it.

Similarly, for the case of even $k$ and  even $w({\bm{m_{1}^{d}}})$ we have
\begin{align}
&\left\lfloor  \frac{w({\bm{m_{1}^{d}}})+w({\bm{m_{s}}})}{2}  \right\rfloor+\left\lfloor  \frac{w({\bm{m_{2}^{d}}})+w({\bm{m_{s}}})}{2}  \right\rfloor\notag\\
&=\left\lfloor \frac{w({\bm{m_{1}^{d}}})}{2}  \right\rfloor+\left\lfloor \frac{w({\bm{m_{s}}})}{2} \right\rfloor+\left\lfloor \frac{w({\bm{m_{2}^{d}}})}{2} \right\rfloor+\left\lfloor  \frac{w({\bm{m_{s}}})}{2} \right\rfloor\notag\\
&=\frac{w({\bm{m_{1}^{d}}})+w({\bm{m_{2}^{d}}})}{2} +2\left\lfloor \frac{w({\bm{m_{s}}})}{2} \right\rfloor\notag\\
&=\frac{k}{2} +2\left\lfloor \frac{w({\bm{m_{s}}})}{2} \right\rfloor.
\end{align}
where, once again, we utilize $w({\bm{m_{1}^{d}}})+w({\bm{m_{2}^{d}}})=k$. A similar procedure can be applied to the case of even $k$ and odd $w({\bm{m_{1}^{d}}})$,  leading to the corresponding result and we omit it.

\emph{The latter}.
Based on the following fact,
\begin{align}
\left\lceil \frac{w({\bm{x}})+w({\bm{y}})}{2} \right\rceil=\left\{\begin{matrix}
 \left\lceil w({\bm{x}})/2 \right\rceil+\left\lceil w({\bm{y}})/2  \right\rceil &, w({\bm{y}})\; {\rm{even}}  \\
  \left\lfloor w({\bm{x}})/2  \right\rfloor+\left\lceil w({\bm{y}})/2  \right\rceil &, w({\bm{y}})\; {\rm{odd}},
\end{matrix}\right.
\end{align}
 analogous procedures can be applied to all the cases of Eq.\,\,(\ref{UL}) leading to the corresponding results and we omit them.
\end{proof}
\section*{APPENDIX C: PROOF OF THE PROPERTIES  }\label{app}
\subsubsection{The proof of  property  1}
\begin{proof}
Through direct calculation, we have
\allowdisplaybreaks[2]
\begin{align}\label{P1}
&\left \| \vec{v}_{0}  \right \|^{2}+\left \| \vec{v}_{1}  \right \|^{2}\notag\\
&=\sum_{{\bm{x_{1} }},{\bm{x_{2} }}\in\left \{ 0,1 \right \}^{N-2 } }(-1)^{\left\lfloor  w({\bm{x_{1}}})/2 \right\rfloor+\left\lfloor  w({\bm{x_{2}}})/2 \right\rfloor}\prod_{i=1}^{N-2} cos\theta _{x_{1},x_{2} }^{i}\notag\\
&\,+ \sum_{{\bm{y_{1} }},{\bm{y_{2} }}\in\left \{ 0,1 \right \}^{N-2 } }(-1)^{\left\lceil  w({\bm{y_{1}}})/2 \right\rceil+\left\lceil  w({\bm{y_{2}}})/2 \right\rceil}\prod_{i=1}^{N-2} cos\theta _{y_{1},y_{2} }^{i},
\end{align}
where throughout the rest of the proof, $\theta _{x_{1},x_{2} }^{i}$ ($\theta _{y_{1},y_{2} }^{i}$) is used to denote the angle between the two measurement directions of
party number $i$ in  sequence ${\bm{x_{1}}}$ and ${\bm{x_{2}}}$ (${\bm{y_{1}}}$ and ${\bm{y_{2}}}$).

Based on the definition of $\left | {\bm{x_{1}}}\bigcap {\bm{x_{2}}}   \right | $ and $\left | {\bm{y_{1}}}\bigcap {\bm{y_{2}}}   \right | $, we can recast Eq.(\ref{P1}) as
\allowdisplaybreaks[2]
\begin{widetext}
\begin{align}
\left \| \vec{v}_{0}  \right \|^{2}+\left \| \vec{v}_{1}  \right \|^{2}&=
2^{N-2}+\sum_{\substack{k=1}}^{N-2} \sum_{\substack{\left | {\bm{x_{1}}}\bigcap {\bm{x_{2}}}   \right |_{j_{1}\cdots j_{k} }=k\\{\bm{x_{1} }},{\bm{x_{2} }}\in \left \{ 0,1 \right \}^{N-2}} }(-1)^{\left\lfloor  w({\bm{x_{1}}})/2 \right\rfloor+\left\lfloor  w({\bm{x_{2}}})/2 \right\rfloor}\prod_{i=1}^{{k} } cos\theta _{x_{1},x_{2} }^{j_{i}}\notag\\
&\quad+2^{N-2}+\sum_{\substack{k=1}}^{N-2} \sum_{\substack{\left | {\bm{y_{1}}}\bigcap {\bm{y_{2}}}   \right |_{j_{1}\cdots j_{k} }=k\\{\bm{y_{1} }},{\bm{y_{2} }}\in \left \{ 0,1 \right \}^{N-2}} }(-1)^{\left\lceil  w({\bm{y_{1}}})/2 \right\rceil+\left\lceil  w({\bm{y_{2}}})/2 \right\rceil}\prod_{i=1}^{{k} } cos\theta _{y_{1},y_{2} }^{j_{i}}\notag\\
&=2^{N-1}\notag\\
&\quad+\sum_{\substack{k=1}}^{N-2} \sum_{\substack{\left | {\bm{x_{1}}}\bigcap {\bm{x_{2}}}   \right |_{j_{1}\cdots j_{k} }=k\\{\bm{x_{1} }},{\bm{x_{2} }}\in \left \{ 0,1 \right \}^{N-2}} }(-1)^{\left\lfloor  w({\bm{x_{1}}})/2 \right\rfloor+\left\lfloor  w({\bm{x_{2}}})/2 \right\rfloor}\prod_{i=1}^{{k} } cos\theta _{x_{1},x_{2} }^{j_{i}}\notag\\
&\quad+\sum_{\substack{k=1}}^{N-2} \sum_{\substack{\left | {\bm{y_{1}}}\bigcap {\bm{y_{2}}}   \right |_{j_{1}\cdots j_{k} }=k\\{\bm{y_{1} }},{\bm{y_{2} }}\in \left \{ 0,1 \right \}^{N-2}} }(-1)^{w({\bm{y_{1}}})+w({\bm{y_{2}}})-\left\lfloor  w({\bm{y_{1}}})/2 \right\rfloor-\left\lfloor  w({\bm{y_{2}}})/2 \right\rfloor}\prod_{i=1}^{{k} } cos\theta _{y_{1},y_{2} }^{j_{i}},
\end{align}
\end{widetext}
where we use the fact $\left\lceil x/2 \right\rceil=x-\left\lfloor x/2 \right\rfloor$ in the last equality.

Using the results of Lemma 3, the above expression can be further simplified as
\allowdisplaybreaks[2]
\begin{widetext}
\begin{align}
&\left \| \vec{v}_{0}  \right \|^{2}+\left \| \vec{v}_{1}  \right \|^{2}=2^{N-1}\notag\\
&\quad+\sum_{\left | {\bm{x_{1}}}\bigcap {\bm{x_{2}}}   \right |=N-2}\sum_{{\bm{x_{1} }},{\bm{x_{2} }}\in \left \{ 0,1 \right \}^{N-2 } }(-1)^{\frac{N-3}{2} }\prod_{i=1}^{N-2} cos\theta _{x_{1},x_{2} }^{i}
\quad+\sum_{\substack{k=odd\\k\neq N-2}} \sum_{\substack{{\bm{x_{1}^{d} }},{\bm{x_{2}^{d} }}\in \left \{ 0,1 \right \}^{k }\\{\bm{x_{s}}}\in \left \{ 0,1 \right \}^{N-2-k}} }(-1)^{\frac{k-1}{2}+ w({\bm{x_{s}}}) }\prod_{i=1}^{{k} } cos\theta _{x_{1}^{d},x_{2}^{d} }^{j_{i}}\notag\\
&\quad+\sum_{\substack{k=even\\k\neq N-2}}\left [  \sum_{\substack{{\bm{x_{1}^{d} }},{\bm{x_{2}^{d} }}\in \left \{ 0,1 \right \}^{k }\\{\bm{x_{s}}}\in \left \{ 0,1 \right \}^{N-2-k}\\w({\bm{x_{1}^{d} }})=odd  } }(-1)^{\frac{k-2}{2}+ 2 \left\lceil w({\bm{x_{s}}})/2 \right\rceil  }\prod_{i=1}^{{k} } cos\theta _{x_{1}^{d},x_{2}^{d} }^{j_{i}}+ \sum_{\substack{{\bm{x_{1}^{d} }},{\bm{x_{2}^{d} }}\in \left \{ 0,1 \right \}^{k }\\{\bm{x_{s}}}\in \left \{ 0,1 \right \}^{N-2-k}\\w({\bm{x_{1}^{d} }})=even  } }(-1)^{\frac{k}{2}+ 2 \left\lfloor w({\bm{x_{s}}})/2 \right\rfloor  }\prod_{i=1}^{{k} } cos\theta _{x_{1}^{d},x_{2}^{d} }^{j_{i}} \right ]\notag\\
&\quad+\sum_{\left | {\bm{y_{1}}}\bigcap {\bm{y_{2}}}   \right |=N-2}\sum_{{\bm{y_{1} }},{\bm{y_{2} }}\in \left \{ 0,1 \right \}^{N-2 } }-(-1)^{\frac{N-3}{2} }\prod_{i=1}^{N-2} cos\theta _{y_{1},y_{2} }^{i}
\quad+\sum_{\substack{k=odd\\k\neq N-2}} \sum_{\substack{{\bm{y_{1}^{d} }},{\bm{y_{2}^{d} }}\in \left \{ 0,1 \right \}^{k }\\{\bm{y_{s}}}\in \left \{ 0,1 \right \}^{N-2-k}} }(-1)^{\frac{3k-1}{2}+3w({\bm{y_{s}}}) }\prod_{i=1}^{{k} } cos\theta _{y_{1}^{d},y_{2}^{d} }^{j_{i}}\notag\\
&\quad+\sum_{\substack{k=even\\k\neq N-2}} \left [\sum_{\substack{{\bm{y_{1}^{d} }},{\bm{y_{2}^{d} }}\in \left \{ 0,1 \right \}^{k }\\{\bm{y_{s}}}\in \left \{ 0,1 \right \}^{N-2-k}\\w({\bm{y_{1}^{d} }})=even  } }(-1)^{\frac{3k}{2}+2w({\bm{y_{s}}})+ 2 \left\lfloor w({\bm{y_{s}}})/2 \right\rfloor  }\prod_{i=1}^{{k} } cos\theta _{y_{1}^{d},y_{2}^{d} }^{j_{i}}\right.\notag\\
&\left. \qquad\qquad\qquad\qquad\qquad\qquad\qquad\qquad\qquad\qquad\quad+ \sum_{\substack{{\bm{y_{1}^{d} }},{\bm{y_{2}^{d} }}\in \left \{ 0,1 \right \}^{k }\\{\bm{y_{s}}}\in \left \{ 0,1 \right \}^{N-2-k}\\w({\bm{y_{1}^{d} }})=odd  } }(-1)^{\frac{3k-2}{2}+2w({\bm{y_{s}}})+ 2 \left\lceil w({\bm{y_{s}}})/2 \right\rceil  }\prod_{i=1}^{{k} } cos\theta _{y_{1}^{d},y_{2}^{d} }^{j_{i}} \right ],
\end{align}
\end{widetext}
where we use the fact that $w({\bm{y_{1}}})+w({\bm{y_{2}}})=k+2w({\bm{y_{s}}})$ for $1\le\left | {\bm{y_{1}}}\bigcap {\bm{y_{2}}}   \right |=k<  N-2$ and $\left\lfloor  w({\bm{y_{1}}})/2 \right\rfloor+\left\lfloor  w({\bm{y_{2}}})/2 \right\rfloor=\frac{N-3}{2} $ for $\left | {\bm{y_{1}}}\bigcap {\bm{y_{2}}}   \right |= N-2 $ with  odd $N$.

By  rearranging the terms we get
\begin{widetext}
\begin{align}\label{P12}
\left \| \vec{v}_{0}  \right \|^{2}+\left \| \vec{v}_{1}  \right \|^{2}
&=2^{N-1}\notag\\
&\quad+\sum_{\substack{k=odd\\k\neq N-2}} \sum_{{\substack{{\bm{x_{1}^{d} }},{\bm{x_{2}^{d} }}\in \left \{ 0,1 \right \}^{k }} }} (-1)^{\frac{k-1}{2}}\prod_{i=1}^{{k} } cos\theta _{x_{1}^{d},x_{2}^{d} }^{j_{i}}\sum_{{\bm{x_{s}}}\in \left \{ 0,1 \right \}^{N-2-k} }(-1)^{ w({\bm{x_{s}}}) }\notag\\
&\quad+\sum_{\substack{k=odd\\k\neq N-2}} \sum_{{\substack{{\bm{y_{1}^{d} }},{\bm{y_{2}^{d} }}\in \left \{ 0,1 \right \}^{k }} }} (-1)^{\frac{3k-1}{2}}\prod_{i=1}^{{k} } cos\theta _{y_{1}^{d},y_{2}^{d} }^{j_{i}}\sum_{{\bm{y_{s}}}\in \left \{ 0,1 \right \}^{N-2-k} }(-1)^{ 3w({\bm{y_{s}}}) }\notag\\
&\quad+\sum_{\substack{k=even\\k\neq N-2}} \sum_{\substack{{\bm{x_{s}}}\in \left \{ 0,1 \right \}^{N-2-k} } }(-1)^{\frac{k}{2}  }\left ( \sum_{\substack{{\bm{x_{1}^{d} }},{\bm{x_{2}^{d} }}\in \left \{ 0,1 \right \}^{k }\\w({\bm{x_{1}^{d} }})=even}}\prod_{i=1}^{{k} } cos\theta _{x_{1}^{d},x_{2}^{d} }^{j_{i}} -\sum_{\substack{{\bm{x_{1}^{d} }},{\bm{x_{2}^{d} }}\in \left \{ 0,1 \right \}^{k }\\w({\bm{x_{1}^{d} }})=odd}}\prod_{i=1}^{{k} } cos\theta _{x_{1}^{d},x_{2}^{d} }^{j_{i}} \right )\notag\\
&\quad+\sum_{\substack{k=even\\k\neq N-2}} \sum_{\substack{{\bm{y_{s}}}\in \left \{ 0,1 \right \}^{N-2-k} } }(-1)^{\frac{3k}{2}  }\left ( \sum_{\substack{{\bm{y_{1}^{d} }},{\bm{y_{2}^{d} }}\in \left \{ 0,1 \right \}^{k }\\w({\bm{y_{1}^{d} }})=even}}\prod_{i=1}^{{k} } cos\theta _{y_{1}^{d},y_{2}^{d} }^{j_{i}} -\sum_{\substack{{\bm{y_{1}^{d} }},{\bm{y_{2}^{d} }}\in \left \{ 0,1 \right \}^{k }\\w({\bm{y_{1}^{d} }})=odd}}\prod_{i=1}^{{k} } cos\theta _{y_{1}^{d},y_{2}^{d} }^{j_{i}} \right )\notag\\
&\quad+
(-1)^{\frac{N-3}{2} }\left ( \sum_{\substack{\left | {\bm{x_{1}}}\bigcap {\bm{x_{2}}}   \right |=N-2\\{\bm{x_{1} }},{\bm{x_{2} }}\in \left \{ 0,1 \right \}^{N-2 }}}\prod_{i=1}^{N-2} cos\theta _{x_{1},x_{2} }^{i}-\sum_{\substack{\left | {\bm{y_{1}}}\bigcap {\bm{y_{2}}}   \right |=N-2\\{\bm{y_{1} }},{\bm{y_{2} }}\in \left \{ 0,1 \right \}^{N-2 }}}\prod_{i=1}^{N-2} cos\theta _{y_{1},y_{2} }^{i}\right ).
\end{align}
\end{widetext}
where we use the fact that even powers of negative one is equal to 1.


Noting that $\prod_{i=1}^{{k} } cos\theta _{x,y }^{i} =\prod_{i=1}^{{k} } \left \langle \vec{a}_{0}^{i}, \vec{a}_{1}^{i}    \right \rangle $ for $\bm{x}\oplus  \bm{y}={\bm{e_{k}}}$, it immediately follows from Lemma 2 that every summation term in Eq. (\ref{P12}) is identically zero, which proves the first property.
\end{proof}

\subsubsection{The proof of  property  2}
\begin{proof}
Noting that  due to ${\bm{x_{1} }},{\bm{x_{2} }}\in {E}^{2 }$, $\left | {\bm{x_{1}}}\bigcap {\bm{x_{2}}}   \right |=1$ is impossible, and similarly for ${\bm{y_{1} }},{\bm{y_{2} }}\in {O}^{2 }$, we have,
\allowdisplaybreaks[2]
\begin{align}\label{Proo2}
&\left \| \vec{u}_{0}  \right \|^{2}+\left \| \vec{u}_{1}  \right \|^{2}=4+\notag\\
&\sum_{\substack{\left | {\bm{x_{1}}}\bigcap {\bm{x_{2}}}   \right |=2 \\{\bm{x_{1} }},{\bm{x_{2} }}\in {E}^{2 }} } (-1)^{\left\lfloor  w({\bm{x_{1}}})/2 \right\rfloor+\left\lfloor  w({\bm{x_{2}}})/2 \right\rfloor}\prod_{i=N-1}^{N} cos\theta _{x_{1},x_{2} }^{i}\notag\\
&\sum_{\substack{\left | {\bm{y_{1}}}\bigcap {\bm{y_{2}}}   \right |=2\\{\bm{y_{1} }},{\bm{y_{2} }}\in {O}^{2 }}}(-1)^{\left\lfloor  w({\bm{y_{1}}})/2 \right\rfloor+\left\lfloor  w({\bm{y_{2}}})/2 \right\rfloor}\prod_{i=N-1}^{N} cos\theta _{y_{1},y_{2} }^{i}\notag\\
&=4+\sum_{\substack{\left | {\bm{x_{1}}}\bigcap {\bm{x_{2}}}   \right |=2  \\{\bm{x_{1} }},{\bm{x_{2} }}\in {E}^{2 }} }(-1)^{1 }\prod_{i=N-1}^{N} cos\theta _{x_{1},x_{2} }^{i}\notag\\
&\qquad+\sum_{\substack{\left | {\bm{y_{1}}}\bigcap {\bm{y_{2}}}   \right |=2  \\{\bm{y_{1} }},{\bm{y_{2} }}\in {O}^{2 }} }(-1)^{0 }\prod_{i=N-1}^{N} cos\theta _{y_{1},y_{2} }^{i}.
\end{align}
where we use the fact that $\left\lfloor  w({\bm{x_{1}}})/2 \right\rfloor+\left\lfloor  w({\bm{x_{2}}})/2 \right\rfloor=1 $ for $\left | {\bm{x_{1}}}\bigcap {\bm{x_{2}}}   \right |= 2$ with ${{\bm{x_{1} }},{\bm{x_{2} }}\in {E}^{2 }}$ and $\left\lfloor  w({\bm{y_{1}}})/2 \right\rfloor+\left\lfloor  w({\bm{y_{2}}})/2 \right\rfloor=0 $ for $\left | {\bm{y_{1}}}\bigcap {\bm{y_{2}}}   \right |= 2$ with ${{\bm{y_{1} }},{\bm{y_{2} }}\in {O}^{2 }}$.

Obviously, the two summation terms in Eq. (\ref{Proo2}) cancel and thus we have proved the second property.
\end{proof}
\subsubsection{The proof of  property  3}
\begin{proof}
Directly applying definition 2 and the results of Lemma 3, we have
\allowdisplaybreaks[2]
\begin{align}
&\vec{v}_{0}\cdot  \vec{v}_{1}
=\notag\\
&\quad\sum_{\substack{\left | {\bm{x_{1}}}\bigcap {\bm{x_{2}}}   \right |=0\\{\bm{x_{1} }},{\bm{x_{2} }}\in \left \{ 0,1 \right \}^{N-2 }}}(-1)^{\left\lfloor  w({\bm{x_{1}}})/2 \right\rfloor+\left\lceil  w({\bm{x_{2}}})/2 \right\rceil}\prod_{i=1}^{N-2} cos\theta _{x_{1},x_{2} }^{i}\notag\\
&+\sum_{\substack{k=even\\k\neq N-2}} \sum_{\substack{{\bm{x_{1}^{d} }},{\bm{x_{2}^{d} }}\in \left \{ 0,1 \right \}^{k }\\{\bm{x_{s}}}\in \left \{ 0,1 \right \}^{N-2-k}} }(-1)^{\frac{k}{2}+ w({\bm{x_{s}}}) }\prod_{i=1}^{{k} } cos\theta _{x_{1}^{d},x_{2}^{d} }^{j_{i}}\notag\\
&+\sum_{\substack{k=odd\\k\neq N-2}}  \sum_{\substack{{\bm{x_{1}^{d} }},{\bm{x_{2}^{d} }}\in \left \{ 0,1 \right \}^{k }\\{\bm{x_{s}}}\in \left \{ 0,1 \right \}^{N-2-k}\\w({\bm{x_{1}^{d} }})=even  } }(-1)^{\frac{k+1}{2}+ 2 \left\lfloor w({\bm{x_{s}}})/2 \right\rfloor  }\prod_{i=1}^{{k} } cos\theta _{x_{1}^{d},x_{2}^{d} }^{j_{i}}\notag\\
&+\sum_{\substack{k=odd\\k\neq N-2}}  \sum_{\substack{{\bm{x_{1}^{d} }},{\bm{x_{2}^{d} }}\in \left \{ 0,1 \right \}^{k }\\{\bm{x_{s}}}\in \left \{ 0,1 \right \}^{N-2-k}\\w({\bm{x_{1}^{d} }})=odd  } }(-1)^{\frac{k-1}{2}+ 2 \left\lceil w({\bm{x_{s}}})/2 \right\rceil  }\prod_{i=1}^{{k} } cos\theta _{x_{1}^{d},x_{2}^{d} }^{j_{i}}\notag\\
&+\sum_{\substack{\left | {\bm{x_{1}}}\bigcap {\bm{x_{2}}}   \right |=N-2\\{\bm{x_{1} }},{\bm{x_{2} }}\in \left \{ 0,1 \right \}^{N-2 }}}(-1)^{\left\lfloor  w({\bm{x_{1}}})/2 \right\rfloor+\left\lceil  w({\bm{x_{2}}})/2 \right\rceil }\prod_{i=1}^{N-2} cos\theta _{x_{1},x_{2} }^{i},
\end{align}
By simplifying and rearranging the terms we get
\allowdisplaybreaks[2]
\begin{widetext}
\begin{align}\label{PP3}
\vec{v}_{0}\cdot  \vec{v}_{1}&=\sum_{{\bm{x }}\in \left \{ 0,1 \right \}^{N-2 } }(-1)^{w({\bm{x})}}\notag\\
&\quad+\sum_{\substack{k=even\\k\neq N-2}} \sum_{{\substack{{\bm{x_{1}^{d} }},{\bm{x_{2}^{d} }}\in \left \{ 0,1 \right \}^{k }} }} (-1)^{\frac{k}{2}}\prod_{i=1}^{{k} } cos\theta _{x_{1}^{d},x_{2}^{d} }^{j_{i}}\sum_{{\bm{x_{s}}}\in \left \{ 0,1 \right \}^{N-2-k} }(-1)^{ w({\bm{x_{s}}}) }\notag\\
&\quad+\sum_{\substack{k=odd\\k\neq N-2}} \sum_{\substack{{\bm{x_{s}}}\in \left \{ 0,1 \right \}^{N-2-k} } }(-1)^{\frac{k-1}{2}  }\left ( \sum_{\substack{{\bm{x_{1}^{d} }},{\bm{x_{2}^{d} }}\in \left \{ 0,1 \right \}^{k }\\w({\bm{x_{1}^{d} }})=odd}}\prod_{i=1}^{{k} } cos\theta _{x_{1}^{d},x_{2}^{d} }^{j_{i}} -\sum_{\substack{{\bm{x_{1}^{d} }},{\bm{x_{2}^{d} }}\in \left \{ 0,1 \right \}^{k }\\w({\bm{x_{1}^{d} }})=even}}\prod_{i=1}^{{k} } cos\theta _{x_{1}^{d},x_{2}^{d} }^{j_{i}} \right )\notag\\
&\quad+(-1)^{\frac{N-3}{2} }\left (  \sum_{\substack{\left | {\bm{x_{1}}}\bigcap {\bm{x_{2}}}   \right |=N-2\\{\bm{x_{1} }},{\bm{x_{2} }}\in \left \{ 0,1 \right \}^{N-2} \\w({\bm{x_{1} }})=even}}\prod_{i=1}^{N-2} cos\theta _{x_{1},x_{2} }^{i}- \sum_{\substack{\left | {\bm{x_{1}}}\bigcap {\bm{x_{2}}}   \right |=N-2\\{\bm{x_{1} }},{\bm{x_{2} }}\in \left \{ 0,1 \right \}^{N-2} \\w({\bm{x_{1} }})=odd}}\prod_{i=1}^{N-2} cos\theta _{x_{1},x_{2} }^{i} \right ).
\end{align}
\end{widetext}
where we use the relation $\left\lceil x/2 \right\rceil=x-\left\lfloor x/2 \right\rfloor$ and employ the following identity for $\left | {\bm{x_{1}}}\bigcap {\bm{x_{2}}}   \right | =N-2$ with ${\bm{x_{1}}},{\bm{x_{2}}}\in \left \{ 0,1 \right \} ^{N-2}$:
\begin{align}
\left\lfloor  \frac{w({\bm{x_{1}}})}{2}  \right\rfloor+\left\lceil  \frac{w({\bm{x_{2}}})}{2}  \right\rceil=\left\{\begin{matrix}
 \frac{N-1}{2}  &, w({\bm{x_{2}}})\; {\rm{odd}} \\
 \frac{N-3}{2}  &, w({\bm{x_{2}}})\; {\rm{even}}.
\end{matrix}\right.
\end{align}
Noting that $\prod_{i=1}^{{k} } cos\theta _{x,y }^{i} =\prod_{i=1}^{{k} } \left \langle \vec{a}_{0}^{i}, \vec{a}_{1}^{i}    \right \rangle $ for $\bm{x}\oplus  \bm{y}={\bm{e_{k}}}$ again, it immediately follows from Lemma 2 that  all the summation terms in Eq. (\ref{PP3}) vanish, which proves the third property.
\end{proof}
\subsubsection{The proof of  property  4}
\begin{proof}
Noting that ${\bm{y_{1} }}\in E^{2 },{\bm{y_{2} }}\in O^{2 }$ means $\left | {\bm{y_{1}}}\bigcap {\bm{y_{2}}}   \right |=1$, thus after  employing again the results of Lemma 3 and Lemma 2 , we obtain
\begin{align}
\vec{u}_{0}\cdot  \vec{u}_{1}&=   \sum_{\substack{{{\bm{y_{1}}}\in {E}^{2}}\\{{\bm{y_{2}}}\in {O}^{2}}}}(-1)^{\left\lfloor  w({\bm{y_{1}}})/2 \right\rfloor+\left\lfloor  w({\bm{y_{2}}})/2 \right\rfloor}\prod_{i=N-1}^{N} cos\theta _{y_{1},y_{2} }^{i}\notag\\
&=\sum_{{\substack{{\bm{y_{1}^{d} }},{\bm{y_{2}^{d} }}\in \left \{ 0,1 \right \}} }}cos\theta _{y_{1}^{d},y_{2}^{d} }^{i}\sum_{{\bm{y_{s}}}\in \left \{ 0,1 \right \} }(-1)^{ w({\bm{m_{s}}}) }\notag\\
&=0
\end{align}
where $i=N-1 $ or $N$.
\end{proof}

\end{document}